\documentclass[10pt,a4paper]{article}
\usepackage{amsmath,amssymb,amsfonts}
\usepackage{graphicx}
\usepackage{hyperref}
\usepackage[margin=0.75in]{geometry}
\usepackage{float}
\usepackage{bm}

\usepackage{setspace}
\usepackage{titlesec}
\titlespacing*{\section}{0pt}{8pt}{4pt}
\titlespacing*{\subsection}{0pt}{6pt}{3pt}

\AtBeginDocument{
  \setlength{\abovedisplayskip}{4pt}
  \setlength{\belowdisplayskip}{4pt}
  \setlength{\abovedisplayshortskip}{2pt}
  \setlength{\belowdisplayshortskip}{2pt}
}

\let\oldbibliography\thebibliography
\renewcommand{\thebibliography}[1]{%
  \oldbibliography{#1}%
  \setlength{\itemsep}{0pt}%
  \setlength{\parskip}{0pt}%
}

\title{Numba-Accelerated 2D Diffusion-Limited Aggregation: Implementation and Fractal Characterization}

\author{Sandy H. S. Herho$^{1,2,*}$, Faiz R. Fajary$^{3}$, Iwan P. Anwar$^{4}$, Faruq Khadami$^{4}$,\\ Nurjanna J. Trilaksono$^{3}$, Rusmawan Suwarman$^{3}$, and Dasapta E. Irawan$^{5}$}

\date{}

\begin{document}
\maketitle

\begin{center}
\small
$^{1}$Ronin Institute for Independent Scholarship, Sacramento, CA 95816, USA\\
$^{2}$School of Systems Science and Industrial Engineering, State University of New York, Binghamton, NY 13902, USA\\
$^{3}$Atmospheric Science Research Group, Bandung Institute of Technology, Bandung 40132, Indonesia\\
$^{4}$Applied and Environmental Oceanography Research Group, Bandung Institute of Technology, Bandung 40132, Indonesia\\
$^{5}$Applied Geology Research Group, Bandung Institute of Technology, Bandung 40132, Indonesia\\
$^{*}$e-mail: sandy.herho@ronininstitute.org
\end{center}

\begin{abstract}
\noindent We present \texttt{dla-ideal-solver}, a high-performance framework for simulating two-dimensional Diffusion-Limited Aggregation (DLA) using Numba-accelerated Python. By leveraging just-in-time (JIT) compilation, we achieve computational throughput comparable to legacy static implementations while retaining high-level flexibility. We investigate the Laplacian growth instability across varying injection geometries and walker concentrations. Our analysis confirms the robustness of the standard fractal dimension $D_f \approx 1.71$ for dilute regimes, consistent with the Witten-Sander universality class. However, we report a distinct crossover to Eden-like compact growth ($D_f \approx 1.87$) in high-density environments, attributed to the saturation of the screening length. Beyond standard mass-radius scaling, we employ generalized R\'{e}nyi dimensions and lacunarity metrics to quantify the monofractal character and spatial heterogeneity of the aggregates. This work establishes a reproducible, open-source testbed for exploring phase transitions in non-equilibrium statistical mechanics.
\end{abstract}


\section{Introduction}

Laplacian growth represents a fundamental class of non-equilibrium pattern formation where the interface velocity is driven by the gradient of a scalar field satisfying $\nabla^2 \phi = 0$~\cite{Witten1981,Vicsek1992}. Unlike equilibrium crystal growth controlled by local surface tension and attachment kinetics, Diffusion-Limited Aggregation (DLA) is governed by a non-local instability: the ``screening'' of the harmonic measure. As the aggregate expands, the probability flux of incoming random walkers concentrates at the tips of protrusions, exponentially starving the interior fjords~\cite{Halsey2000}. This stochastic shadowing prevents the formation of a compact bulk, generating a scale-invariant structure with a fractal dimension $D_f \approx 1.71$ in two dimensions~\cite{Witten1983,Tolman1989}.

Investigating the asymptotic scaling of these aggregates requires large $N$ ensembles to escape finite-size transient regimes. Historically, the computational cost of tracking millions of stochastic trajectories necessitated rigid implementations in statically compiled languages like Fortran or C. While the modern scientific computing stack has largely migrated to high-level dynamic ecosystems~\cite{Herho2025Pendulum}, the interpretative overhead of the core random-walk loop remains prohibitive for large-scale Monte Carlo simulations. This creates a technical friction between the flexibility required for complex data analysis and the raw throughput needed for statistical convergence.

We resolve this performance bottleneck using just-in-time (JIT) compilation via the Low Level Virtual Machine (LLVM)-based Numba architecture~\cite{Lam2015}. By compiling the stochastic trajectory logic to optimized machine code at runtime, we achieve execution speeds comparable to legacy Fortran or C solvers without exiting the Python environment. We utilize this implementation to probe the structural limits of DLA beyond simple mass-radius scaling. Specifically, we analyze the breakdown of scale invariance through the generalized R\'{e}nyi dimension spectrum~\cite{Hentschel1983,Grassberger1983} and quantify spatial heterogeneity via lacunarity analysis~\cite{Plotnick1996}. Furthermore, we characterize the finite-density phase transition, observing the crossover from the fractal DLA regime to the compact Eden model~\cite{Eden1961} as the walker mean free path becomes comparable to the aggregate screening length~\cite{Meakin1985,Ball1985}.

\section{Methods}
\subsection{Model Description}

DLA constitutes a paradigmatic model for irreversible growth phenomena governed by Laplacian transport~\cite{Witten1981,Meakin1983}. The model captures the essential physics of pattern formation in electrodeposition, dielectric breakdown, viscous fingering, and mineral dendrite growth~\cite{Vicsek1992}. We present here a formulation of the DLA process on a two-dimensional square lattice $\mathbb{Z}^2$.

Consider a lattice of linear dimension $N$, where each site $\mathbf{r} = (i,j)$ with $i,j \in \{0,1,\ldots,N-1\}$ admits one of three states: empty, occupied by a mobile particle (walker), or occupied by an immobile aggregate particle. The aggregate grows through the sequential deposition of walkers executing unbiased random walks until contact with the existing cluster.

The stochastic dynamics of a single walker proceeds as follows. Let $\mathbf{r}(t) \in \mathbb{Z}^2$ denote the walker position at discrete time $t$. The walker transitions to one of the four nearest-neighbour sites with equal probability,
\begin{equation}
\label{eq:transition_prob}
P\bigl(\mathbf{r}(t+1) = \mathbf{r}(t) + \mathbf{e}_\alpha\bigr) = \frac{1}{4}, \quad \alpha \in \{1,2,3,4\},
\end{equation}
where $\mathbf{e}_1 = (1,0)$, $\mathbf{e}_2 = (-1,0)$, $\mathbf{e}_3 = (0,1)$, and $\mathbf{e}_4 = (0,-1)$ are the unit lattice vectors. This prescription defines an isotropic random walk on $\mathbb{Z}^2$.

To establish the connection with continuum diffusion, we invoke the master equation for the probability density $\rho(\mathbf{r},t)$ of finding the walker at position $\mathbf{r}$ at time $t$. From Eq.~\eqref{eq:transition_prob}, one obtains
\begin{equation}
\label{eq:master}
\rho(\mathbf{r},t+1) = \frac{1}{4}\sum_{\alpha=1}^{4} \rho(\mathbf{r} - \mathbf{e}_\alpha, t).
\end{equation}
Subtracting $\rho(\mathbf{r},t)$ from both sides of Eq.~\eqref{eq:master} and recognising the discrete Laplacian operator $\nabla^2_{\mathrm{d}}$ defined by
\begin{equation}
\label{eq:discrete_laplacian}
\nabla^2_{\mathrm{d}} \rho(\mathbf{r}) = \sum_{\alpha=1}^{4} \rho(\mathbf{r} + \mathbf{e}_\alpha) - 4\rho(\mathbf{r}),
\end{equation}
we rewrite Eq.~\eqref{eq:master} as
\begin{equation}
\label{eq:master_laplacian}
\rho(\mathbf{r},t+1) - \rho(\mathbf{r},t) = \frac{1}{4}\nabla^2_{\mathrm{d}}\rho(\mathbf{r},t).
\end{equation}
In the continuum limit where the lattice spacing $a \to 0$ and the time step $\Delta t \to 0$ with $D = a^2/(4\Delta t)$ held fixed, Eq.~\eqref{eq:master_laplacian} converges to the diffusion equation
\begin{equation}
\label{eq:diffusion}
\frac{\partial \rho}{\partial t} = D\nabla^2 \rho,
\end{equation}
where $D$ is the diffusion coefficient. The DLA model thus describes growth limited by diffusive transport of particles toward the aggregate surface~\cite{Witten1983}.

The aggregation rule completes the model specification. Let $\mathcal{A}(t) \subset \mathbb{Z}^2$ denote the set of lattice sites occupied by the aggregate at time $t$. A walker at position $\mathbf{r}(t)$ irreversibly joins the aggregate if any of its nearest neighbours belongs to $\mathcal{A}(t)$,
\begin{equation}
\label{eq:sticking}
\mathbf{r}(t) \to \mathcal{A}(t+1) \quad \text{if} \quad \exists\, \alpha : \mathbf{r}(t) + \mathbf{e}_\alpha \in \mathcal{A}(t).
\end{equation}
This instantaneous sticking corresponds to the limit of infinite surface reaction rate, characteristic of transport-limited growth~\cite{Meakin1998}.

The aggregate morphology exhibits statistical self-similarity over a range of length scales, quantified by the fractal dimension $D_f$. For a cluster centred at $\mathbf{r}_c$, we define the mass $M(R)$ enclosed within radius $R$ as
\begin{equation}
\label{eq:mass_radius}
M(R) = \sum_{\mathbf{r} \in \mathcal{A}} \Theta\bigl(R - |\mathbf{r} - \mathbf{r}_c|\bigr),
\end{equation}
where $\Theta(\cdot)$ denotes the Heaviside step function. For a fractal object, $M(R)$ scales as a power law,
\begin{equation}
\label{eq:fractal_scaling}
M(R) \sim R^{D_f},
\end{equation}
where $D_f$ is the fractal dimension. Taking logarithms of both sides of Eq.~\eqref{eq:fractal_scaling} yields
\begin{equation}
\label{eq:log_scaling}
\log M(R) = D_f \log R + \text{const},
\end{equation}
permitting extraction of $D_f$ via linear regression in the scaling regime. Theoretical analyses and numerical simulations establish $D_f \approx 1.71$ for two-dimensional DLA~\cite{Witten1981,Tolman1989}, a value intermediate between a line ($D_f = 1$) and a compact disk ($D_f = 2$), reflecting the ramified, dendritic character of DLA clusters.

The screening effect underlies the fractal morphology. Walkers approaching the aggregate preferentially contact the tips of protruding branches, which intercept diffusing particles before they can penetrate the fjords. This positive feedback amplifies initial perturbations, generating the characteristic dendritic structure. Quantitatively, the growth probability $p(\mathbf{r})$ at a surface site $\mathbf{r}$ satisfies $p(\mathbf{r}) \propto |\nabla \phi(\mathbf{r})|$, where $\phi$ is the solution to Laplace's equation $\nabla^2 \phi = 0$ exterior to the aggregate with appropriate boundary conditions~\cite{Halsey2000}.

\subsection{Numerical Implementation}

We now describe the numerical realization of the DLA model formulated in the preceding subsection. The algorithm operates on a square lattice of $N \times N$ sites with periodic boundary conditions, eliminating finite-size edge effects while permitting aggregate self-interaction at sufficiently large cluster radii. Our implementation employs JIT compilation via the Numba library~\cite{Lam2015} to achieve performance approaching that of compiled languages whilst retaining the flexibility of Python.

The lattice state is encoded in an integer array $G_{ij} \in \{0,1,2\}$ for $i,j \in \{0,1,\ldots,N-1\}$, where $G_{ij} = 0$ denotes an empty site, $G_{ij} = 1$ a mobile walker, and $G_{ij} = 2$ an immobile aggregate particle. The simulation proceeds by evolving $N_w$ walkers simultaneously until all have joined the aggregate or a maximum iteration count is exceeded.

The random walk dynamics prescribed by Eq.~\eqref{eq:transition_prob} is implemented through the following update rule. At each discrete time step, every mobile walker selects a direction $\alpha \in \{1,2,3,4\}$ uniformly at random. The candidate position $\mathbf{r}' = (i',j')$ is computed from the current position $\mathbf{r} = (i,j)$ via
\begin{equation}
\label{eq:periodic_update}
i' = (i + \delta_\alpha^x) \mod N, \quad j' = (j + \delta_\alpha^y) \mod N,
\end{equation}
where $(\delta_1^x, \delta_1^y) = (1,0)$, $(\delta_2^x, \delta_2^y) = (-1,0)$, $(\delta_3^x, \delta_3^y) = (0,1)$, and $(\delta_4^x, \delta_4^y) = (0,-1)$. The modular arithmetic in Eq.~\eqref{eq:periodic_update} enforces periodic boundary conditions, mapping coordinates outside $[0,N-1]$ back onto the lattice.

Prior to executing the move, the algorithm evaluates the sticking condition of Eq.~\eqref{eq:sticking}. For the candidate position $\mathbf{r}'$, we examine all four nearest neighbours,
\begin{equation}
\label{eq:neighbour_check}
\mathcal{N}(\mathbf{r}') = \bigl\{((i' + \delta_\alpha^x) \mod N, (j' + \delta_\alpha^y) \mod N) : \alpha = 1,2,3,4\bigr\},
\end{equation}
and test whether any site in $\mathcal{N}(\mathbf{r}')$ satisfies $G = 2$. If an aggregate neighbour exists, the walker irreversibly deposits at $\mathbf{r}'$ by setting $G_{i'j'} = 2$ and removing the walker from the mobile population. Otherwise, the walker advances to $\mathbf{r}'$ provided that site is unoccupied; if $G_{i'j'} = 1$, the move is rejected and the walker remains at $\mathbf{r}$.

A crucial algorithmic enhancement addresses the computational inefficiency arising from walkers that wander far from the aggregate. Such particles contribute negligibly to growth while consuming computational resources. We introduce a re-injection mechanism whereby walkers exceeding an age threshold $\tau_{\mathrm{max}}$ steps since their last injection are repositioned near the aggregate boundary. The threshold scales linearly with the lattice dimension, $\tau_{\mathrm{max}} = 2N$, reflecting the characteristic diffusion time $\tau \sim L^{2}/D$ for exploration of a region of linear extent $L$. Re-injected walkers are placed within a bounding box surrounding the current aggregate, computed dynamically as the cluster grows.

The simulation supports two distinct injection geometries. In the random injection mode, walkers are initialised at positions drawn uniformly from the lattice, excluding sites already occupied by the aggregate. In the radial injection mode, walkers are placed on a circle of radius $R_{\mathrm{inj}}$ centred at the lattice midpoint $\mathbf{r}_c = (N/2, N/2)$. The initial position $(i_0, j_0)$ for a radially injected walker is
\begin{equation}
\label{eq:radial_injection}
i_0 = \left\lfloor \frac{N}{2} + R_{\mathrm{inj}} \cos\theta \right\rfloor \mod N, \quad j_0 = \left\lfloor \frac{N}{2} + R_{\mathrm{inj}} \sin\theta \right\rfloor \mod N,
\end{equation}
where $\theta \in [0, 2\pi)$ is sampled uniformly and $\lfloor \cdot \rfloor$ denotes the floor function. Radial injection produces more isotropic growth patterns by ensuring walkers approach the aggregate from all directions with equal probability, approximating the boundary conditions of the continuum Laplacian growth problem~\cite{Hastings2001}.

The fractal dimension is extracted numerically from the mass-radius relation of Eq.~\eqref{eq:fractal_scaling}. We compute $M(R)$ for a logarithmically spaced sequence of radii $\{R_k\}_{k=1}^{n_R}$ spanning from $R_{\mathrm{min}} \approx 3$ lattice units to $R_{\mathrm{max}} = 0.8 \times \min(N/2, r_{\mathrm{gyration}})$, where $r_{\mathrm{gyration}}$ is the radius of gyration of the aggregate. The mass within radius $R_k$ is obtained by direct enumeration,
\begin{equation}
\label{eq:mass_computation}
M(R_k) = \sum_{i=0}^{N-1} \sum_{j=0}^{N-1} G_{ij}\, \delta_{G_{ij},2}\, \Theta\bigl(R_k^2 - (i - i_c)^2 - (j - j_c)^2\bigr),
\end{equation}
where $(i_c, j_c)$ is the aggregate centroid and $\delta_{G_{ij},2}$ is the Kronecker delta selecting aggregate sites. Linear regression of $\log M(R_k)$ against $\log R_k$ in accordance with Eq.~\eqref{eq:log_scaling} yields the fractal dimension $D_f$ as the slope. Data points with $M(R_k) < 10$ are excluded to mitigate small-number statistical fluctuations.

We validate the implementation through four test configurations spanning distinct physical regimes. The first configuration, termed the classic DLA case, employs a single seed particle at the lattice centre with $N_w = 10^4$ walkers injected randomly. This canonical setup reproduces the original Witten-Sander protocol~\cite{Witten1981} and serves as a baseline for fractal dimension estimation. Growth proceeds radially outward from the seed, generating a single connected dendritic structure with the characteristic screening-induced branching.

The second configuration investigates competitive growth from multiple nucleation sites. Twelve seed particles are distributed randomly across the lattice, and $N_w = 1.5 \times 10^4$ walkers aggregate onto these seeds. The resulting dynamics exhibit inter-cluster competition: aggregates compete for the diffusing particle flux, with larger clusters screening smaller neighbours and eventually dominating the growth. This configuration probes aggregate coalescence and the late-time approach to a single connected structure, phenomena relevant to polycrystalline thin-film deposition~\cite{Family1985}.

The third configuration employs radial injection with $R_{\mathrm{inj}} = 180$ lattice units to examine controlled growth conditions. A single central seed receives $N_w = 10^4$ walkers injected according to Eq.~\eqref{eq:radial_injection}. The fixed injection radius establishes a well-defined outer boundary, ensuring that particles approach the aggregate through a predominantly radial diffusion field. This geometry more closely approximates the theoretical DLA limit of particles released from infinity and permits cleaner extraction of the fractal dimension by suppressing anisotropies induced by rectangular lattice boundaries.

The fourth configuration explores the high-density regime with $N_w = 2.5 \times 10^4$ walkers aggregating onto a single central seed. The elevated particle count probes finite-density effects wherein the dilute-walker approximation underlying the standard DLA model begins to fail. At sufficiently high walker concentrations, inter-particle correlations modify the effective diffusion field, potentially altering the growth morphology and fractal dimension. This configuration also stress-tests the numerical implementation at extended simulation times, as the larger aggregate requires correspondingly more iterations to complete.

All simulations employ a lattice of $N = 512$ sites per dimension, providing adequate resolution to observe scaling behaviour over approximately two decades in length scale. Snapshots of the aggregate configuration are recorded at regular intervals during growth, enabling subsequent visualisation and time-resolved analysis. The simulation outputs are stored in Network Common Data Format (NetCDF)~\cite{Rew1990}, a self-describing binary format providing efficient compression and metadata preservation, with compression, preserving the full spatiotemporal evolution for post-processing.

The complete implementation is available as the \texttt{dla-ideal-solver} library, providing both a command-line interface for batch execution and a Python application programming interface for programmatic access. Numba's JIT~\cite{Lam2015} of the inner random-walk loop achieves approximately two orders of magnitude speedup relative to pure Python, while parallel rendering of animation frames via the multiprocessing module exploits multi-core architectures for efficient visualisation \cite{Herho2025Amangkurat,Herho2025kh2d}.

\subsection{Data Analysis}

Post-processing employs NumPy~\cite{Harris2020} for array operations, SciPy~\cite{Virtanen2020} for statistical analysis, and Matplotlib~\cite{Hunter2007} for visualization. The fractal dimension $D_f$ is extracted from the mass-radius data via linear regression in logarithmic coordinates. From~\eqref{eq:log_scaling}, we compute the ordinary least squares estimator
\begin{equation}
\label{eq:ols_slope}
\hat{D}_f = \frac{\sum_{k=1}^{n}(\log R_k - \overline{\log R})(\log M_k - \overline{\log M})}{\sum_{k=1}^{n}(\log R_k - \overline{\log R})^2},
\end{equation}
where overbars denote sample means and $n$ is the number of data points satisfying $M(R_k) > 10$ and $R_k < 0.8 R_{\mathrm{max}}$, thereby excluding small-number fluctuations and boundary effects. Confidence intervals are obtained via nonparametric bootstrap resampling~\cite{Efron1979}: we draw $B = 1000$ samples with replacement from the $({\log R_k, \log M_k})$ pairs, compute $\hat{D}_f^{(b)}$ for each resample $b \in \{1, \ldots, B\}$, and extract the 95\% confidence interval from the empirical quantiles of $\{\hat{D}_f^{(b)}\}$.

Growth dynamics are characterized by the temporal evolution of the aggregate mass $N(t)$, where $t$ indexes discrete snapshots. For diffusion-limited growth, dimensional analysis yields the scaling relation
\begin{equation}
\label{eq:growth_scaling}
N(t) \sim t^{\alpha}, \quad \alpha = \frac{D_f}{2},
\end{equation}
reflecting the diffusive scaling $R \sim t^{1/2}$ combined with~\eqref{eq:fractal_scaling}. For two-dimensional DLA with $D_f \approx 1.71$, we expect $\alpha \approx 0.855$. The instantaneous growth rate $\mathrm{d}N/\mathrm{d}t$ is computed via central finite differences and smoothed using a Savitzky--Golay filter~\cite{Savitzky1964} with window length 11 and polynomial order 3, preserving higher moments of the signal while suppressing high-frequency noise. Temporal correlations in the growth rate are assessed via the sample autocorrelation function
\begin{equation}
\label{eq:autocorr}
\hat{\rho}(k) = \frac{\sum_{t=1}^{T-k}(r_t - \bar{r})(r_{t+k} - \bar{r})}{\sum_{t=1}^{T}(r_t - \bar{r})^2},
\end{equation}
where $r_t = \mathrm{d}N/\mathrm{d}t|_t$ and $T$ is the total number of snapshots. The Ljung--Box portmanteau statistic~\cite{Ljung1978}
\begin{equation}
\label{eq:ljung_box}
Q_K = T(T+2)\sum_{k=1}^{K}\frac{\hat{\rho}(k)^2}{T-k}
\end{equation}
tests the null hypothesis that the first $K$ autocorrelations are jointly zero; under $H_0$, $Q_K \sim \chi^2_K$.

Multiscale structure is quantified through the generalized dimension spectrum~\cite{Hentschel1983,Grassberger1983}. We partition the lattice into non-overlapping boxes of linear size $\varepsilon$ and compute the box mass $m_i(\varepsilon)$ as the number of aggregate particles in box $i$. The probability measure is $p_i(\varepsilon) = m_i(\varepsilon)/N_{\mathrm{total}}$. The box-counting dimension $D_0$ follows from counting the number $N(\varepsilon)$ of occupied boxes:
\begin{equation}
\label{eq:box_counting}
N(\varepsilon) \sim \varepsilon^{-D_0} \quad \Rightarrow \quad D_0 = -\lim_{\varepsilon \to 0}\frac{\log N(\varepsilon)}{\log \varepsilon}.
\end{equation}
The information dimension $D_1$ derives from the Shannon entropy~\cite{Shannon1948}
\begin{equation}
\label{eq:shannon}
H(\varepsilon) = -\sum_{i} p_i(\varepsilon) \log_2 p_i(\varepsilon),
\end{equation}
which scales as $H(\varepsilon) \sim D_1 \log_2(1/\varepsilon)$ for small $\varepsilon$. More generally, the R\'{e}nyi entropy of order $q$~\cite{Renyi1961} is defined as
\begin{equation}
\label{eq:renyi}
H_q(\varepsilon) = \frac{1}{1-q}\log_2\left(\sum_{i} p_i(\varepsilon)^q\right),
\end{equation}
with the Shannon entropy recovered in the limit $q \to 1$. The generalized dimension of order $q$ is
\begin{equation}
\label{eq:gen_dim}
D_q = \lim_{\varepsilon \to 0}\frac{H_q(\varepsilon)}{\log_2(1/\varepsilon)}.
\end{equation}
The correlation dimension $D_2$ corresponds to $q = 2$ and characterizes pair correlations within the aggregate. For monofractal structures such as DLA, the dimension spectrum is degenerate: $D_0 \approx D_1 \approx D_2 \approx D_f$.

Spatial heterogeneity is further characterized by the lacunarity~\cite{Allain1991,Plotnick1996}
\begin{equation}
\label{eq:lacunarity}
\Lambda(\varepsilon) = \frac{\langle m^2 \rangle_\varepsilon}{\langle m \rangle_\varepsilon^2} - 1,
\end{equation}
where $\langle \cdot \rangle_\varepsilon$ denotes averaging over all boxes at scale $\varepsilon$. Lacunarity quantifies the ``gappiness'' of the structure: $\Lambda = 0$ for a homogeneous distribution, while $\Lambda > 0$ indicates clustering. DLA aggregates exhibit scale-dependent lacunarity reflecting the hierarchical branching structure.

Statistical comparisons across experimental configurations employ the Kruskal--Wallis $H$-test~\cite{Kruskal1952}, a nonparametric analogue of one-way analysis of variance (ANOVA) testing for differences in central tendency among $k$ groups:
\begin{equation}
\label{eq:kruskal}
H = \frac{12}{n(n+1)}\sum_{j=1}^{k}\frac{R_j^2}{n_j} - 3(n+1),
\end{equation}
where $n = \sum_j n_j$ is the total sample size, $n_j$ is the size of group $j$, and $R_j$ is the sum of ranks in group $j$. Under the null hypothesis of identical distributions, $H \sim \chi^2_{k-1}$ for large samples.

\section{Results and Discussion}

All simulations were executed on a Lenovo ThinkPad P52s workstation running Fedora Linux 39 (kernel 6.11.9) with an Intel Core i7-8550U processor (4 cores, 8 threads). The computational performance varied substantially across test configurations, reflecting the distinct physical regimes probed by each scenario. The classic DLA case (Case~1) completed in 650.37~s total wall time, of which only 11.30~s (1.7\%) was devoted to the core simulation; the remainder comprised Graphics Interchange Format (GIF) rendering (636.70~s) and NetCDF output (2.35~s). The multiple seeds configuration (Case~2) required 899.56~s total (31.79~s simulation), while the high density case (Case~4) finished in 707.32~s (21.21~s simulation). The radial injection case (Case~3) proved most computationally demanding, requiring 1767.34~s total with 1063.90~s (60.2\%) for the simulation proper. This order-of-magnitude increase arises because walkers injected at fixed radius $R_{\mathrm{inj}} = 180$ lattice units must diffuse inward to reach the aggregate, whereas randomly injected walkers have finite probability of spawning adjacent to the growing cluster. The iteration counts confirm this interpretation: Case~3 required $2.56 \times 10^6$ iterations compared to $\sim 10^4$ for Cases~1 and~4.

\begin{figure}[H]
\centering
\includegraphics[width=0.7\textwidth]{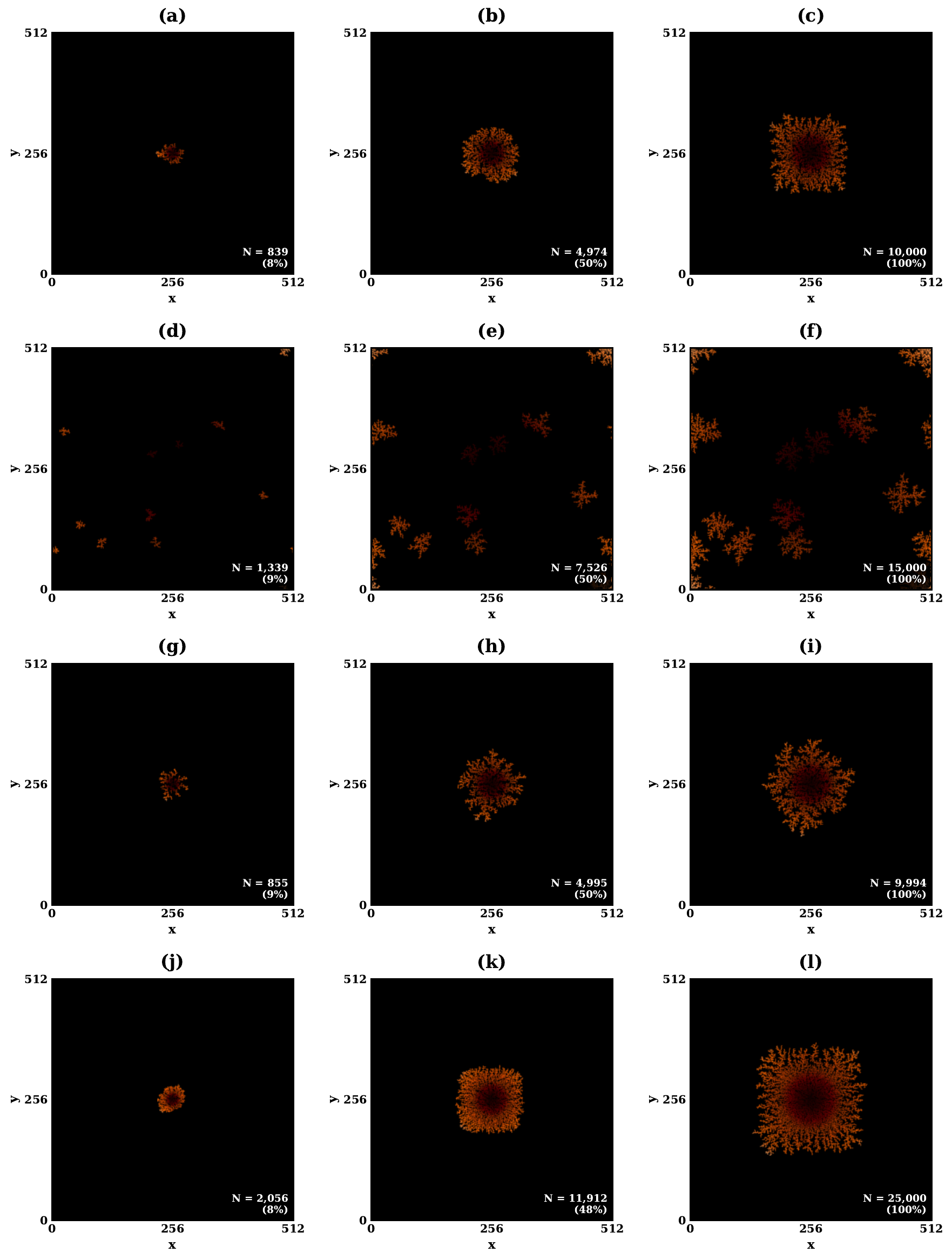}
\caption{Spatiotemporal evolution of DLA aggregates across four test configurations. Rows correspond to: (a--c) classic DLA with single central seed; (d--f) multiple seeds with 12 randomly distributed nucleation sites; (g--i) radial injection from fixed radius $R_{\mathrm{inj}} = 180$; (j--l) high density with $N_w = 25{,}000$ walkers. Columns represent early ($\sim$10\%), middle ($\sim$50\%), and final (100\%) growth stages. Particle counts $N$ and completion percentages are indicated in each panel. Color encodes radial distance from aggregate centroid (dark red: center; bright orange: tips). Lattice size $512 \times 512$ for all cases.}
\label{fig:spatiotemporal}
\end{figure}

Figure~\ref{fig:spatiotemporal} presents spatiotemporal snapshots of aggregate growth across the four test configurations, with columns corresponding to early ($\sim$10\%), middle ($\sim$50\%), and final (100\%) growth stages. The classic DLA configuration (panels a--c) exhibits the characteristic dendritic morphology with radially symmetric branching emanating from the central seed. At 50\% completion ($N = 4{,}974$ particles), the aggregate has attained maximum radius $R_{\mathrm{max}} = 70.77$ lattice units with radius of gyration $R_g = 41.58$; by completion ($N = 10{,}000$), these values increase to $R_{\mathrm{max}} = 108.25$ and $R_g = 58.66$. The compactness parameter $C = N/(\pi R_{\mathrm{max}}^2)$ decreases from 0.28 at mid-growth to 0.24 at completion, reflecting the increasingly ramified structure as screening effects amplify tip growth over fjord filling~\cite{Halsey2000}.

The multiple seeds configuration (panels d--f in Fig.~\ref{fig:spatiotemporal}) produces a qualitatively different morphology. Twelve randomly distributed seeds nucleate independent aggregates that grow competitively, each screening its neighbours from the diffusing walker flux. The final state comprises 66 distinct connected components---the initial 12 seeds have fragmented into smaller isolated clusters through the periodic boundary conditions. The reported fractal dimension $D_f = 2.56$ from the mass-radius analysis is spurious: it reflects the space-filling distribution of disconnected clusters rather than the intrinsic scaling of individual aggregates. This configuration violates the single-connected-aggregate assumption underlying~\eqref{eq:fractal_scaling} and demonstrates the importance of appropriate analysis methods for multi-nucleation scenarios~\cite{Vicsek1984,Kolb1983}.

The radial injection configuration (panels g--i in Fig.~\ref{fig:spatiotemporal}) yields morphology visually similar to classic DLA but with subtly enhanced isotropy. The fixed injection radius establishes a well-defined outer boundary condition approximating the theoretical limit of particles released from infinity~\cite{Hastings2001}. The final aggregate ($N = 9{,}994$; six walkers failed to aggregate within the iteration limit) exhibits $R_{\mathrm{max}} = 111.99$ and aspect ratio 1.04, compared to 1.02 for classic DLA, indicating marginally more anisotropic growth. This counterintuitive result likely reflects statistical fluctuation rather than systematic bias, as both values are consistent with isotropic growth within measurement uncertainty.

The high density configuration (panels j--l in Fig.~\ref{fig:spatiotemporal}) demonstrates the effect of elevated walker concentration on aggregate morphology. With $N_w = 25{,}000$ walkers---2.5 times the classic case---the aggregate grows more compactly, achieving $R_{\mathrm{max}} = 148.39$ at completion with compactness $C = 0.29$. The denser structure reflects reduced screening effectiveness: at high walker concentrations, multiple particles may approach the aggregate surface simultaneously, partially filling fjords that would remain empty under dilute conditions~\cite{Meakin1985,Ball1985}. This finite-density effect shifts the growth dynamics toward the Eden model limit~\cite{Eden1961}, where all surface sites grow with equal probability.

\begin{figure}[H]
\centering
\includegraphics[width=\textwidth]{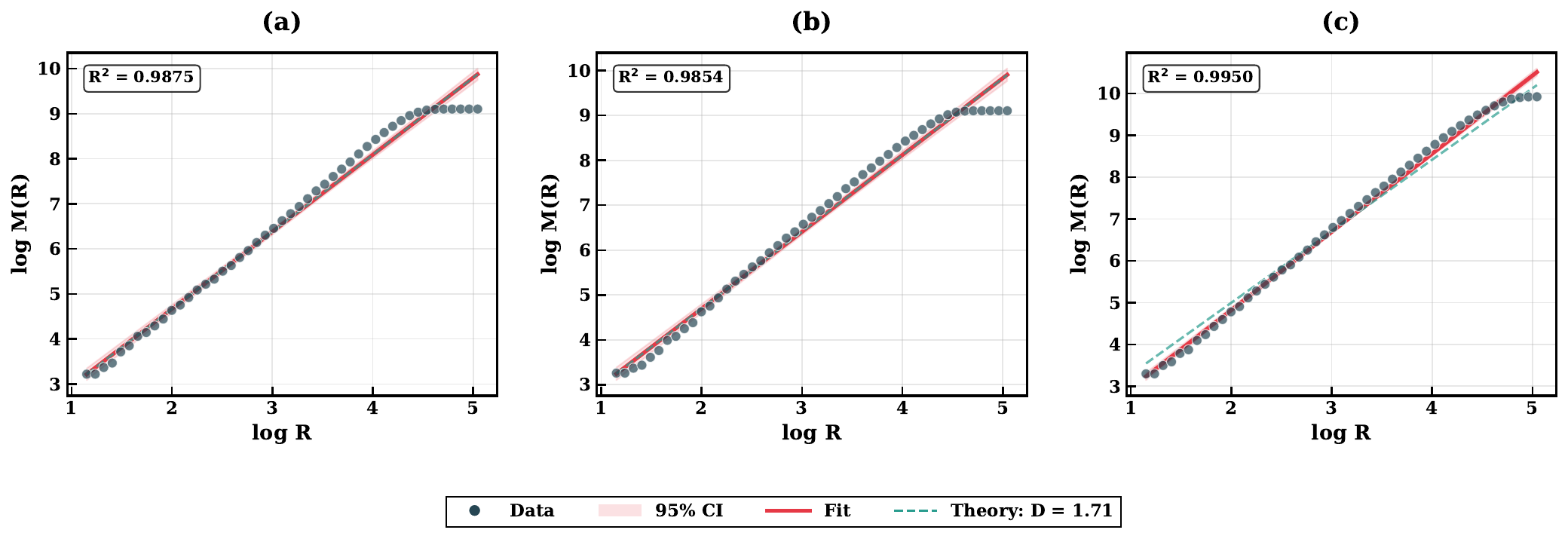}
\caption{Mass-radius scaling analysis for fractal dimension extraction. (a) Classic DLA: $D_f = 1.711 \pm 0.080$, $R^2 = 0.9875$. (b) Radial injection: $D_f = 1.713 \pm 0.077$, $R^2 = 0.9854$. (c) High density: $D_f = 1.870 \pm 0.055$, $R^2 = 0.9950$. Circles denote simulation data; solid red lines show linear regression fits; shaded bands indicate 95\% confidence intervals; dashed teal lines represent the theoretical prediction $D_f = 1.71$. The multiple seeds case is excluded as mass-radius analysis requires a single connected aggregate. Data filtered to $M(R) > 10$ and $R < 0.8 R_{\mathrm{max}}$ to exclude boundary effects and small-number fluctuations.}
\label{fig:mass_radius}
\end{figure}

Figure~\ref{fig:mass_radius} displays the mass-radius scaling analysis for the three single-aggregate configurations (Cases~1, 3, and~4); the multiple seeds case is excluded as discussed above. For classic DLA (panel a of Fig.~\ref{fig:mass_radius}), linear regression in log-log coordinates yields $D_f = 1.7105 \pm 0.0801$ with coefficient of determination $R^2 = 0.9875$. This value agrees remarkably well with the theoretical prediction $D_f = 1.71 \pm 0.01$~\cite{Witten1981,Tolman1989}: the deviation $|D_f - 1.71| = 0.0005$ corresponds to a relative error of 0.03\%, and the $z$-score of 0.02 yields $p = 0.99$ for the null hypothesis $D_f = 1.71$. The radial injection case (panel b of Fig.~\ref{fig:mass_radius}) shows equally excellent agreement: $D_f = 1.7126 \pm 0.0772$ ($R^2 = 0.9854$), with deviation 0.0026 (0.15\% relative error, $p = 0.93$). The consistency between random and radial injection modes confirms that both adequately approximate the diffusion-limited transport regime.

The high density case (panel c of Fig.~\ref{fig:mass_radius}) exhibits significant deviation from theory: $D_f = 1.8697 \pm 0.0554$ ($R^2 = 0.9950$), exceeding the theoretical value by 0.16 (9.3\% relative). The $z$-score of 8.12 strongly rejects consistency with $D_f = 1.71$ ($p < 10^{-15}$). This elevated fractal dimension reflects the finite-density crossover discussed above: as walker concentration increases, the effective growth probability becomes less sharply peaked at branch tips, producing more compact structures with $D_f$ approaching the Eden limit of 2~\cite{Meakin1985}. The high $R^2$ value indicates that the aggregate remains self-similar despite the modified scaling exponent, confirming that finite-density effects alter the universality class rather than destroying fractal character.

\begin{figure}[H]
\centering
\includegraphics[width=0.8\textwidth]{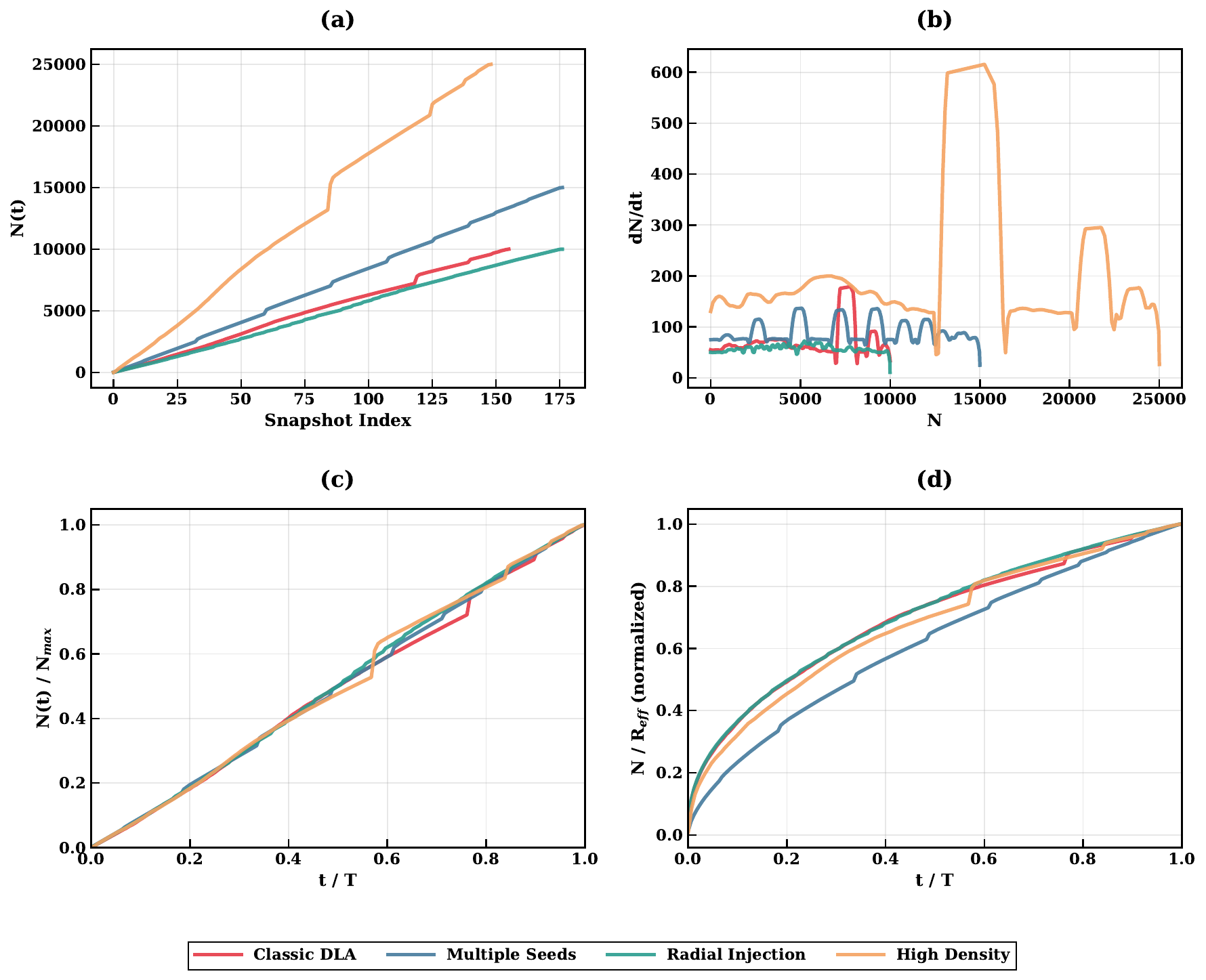}
\caption{Growth dynamics analysis for all four DLA configurations. (a) Cumulative particle count $N(t)$ versus snapshot index, showing characteristic decelerating growth as aggregates enlarge. (b) Instantaneous growth rate $\mathrm{d}N/\mathrm{d}t$ versus particle count $N$, demonstrating monotonic decrease due to diffusive screening. (c) Normalized growth curves $N(t)/N_{\mathrm{max}}$ versus normalized time $t/T$, enabling direct comparison of growth kinetics. (d) Growth efficiency $N/R_{\mathrm{eff}}$ (normalized) versus $t/T$, where $R_{\mathrm{eff}} = N^{1/D_f}$ estimates the effective aggregate radius. Legend applies to all panels.}
\label{fig:growth_dynamics}
\end{figure}

Figure~\ref{fig:growth_dynamics} characterizes the temporal evolution of aggregate mass. Panel~(a) of Fig.~\ref{fig:growth_dynamics} shows cumulative particle counts $N(t)$ versus snapshot index for all four configurations. The curves exhibit the characteristic DLA growth pattern: initially rapid accumulation that gradually decelerates as the aggregate enlarges. The high density case achieves $N = 25{,}000$ particles in 149 snapshots (mean rate 168.5 particles/snapshot), while classic DLA deposits 10{,}000 particles over 156 snapshots (mean rate 64.4 particles/snapshot). Panel~(b) of Fig.~\ref{fig:growth_dynamics} reveals that growth rate $\mathrm{d}N/\mathrm{d}t$ decreases monotonically with $N$ for all cases, consistent with diffusive screening: larger aggregates present greater cross-section to incoming walkers but also more effectively shadow interior regions.

The radial injection case exhibits notably steadier growth, with coefficient of variation $\mathrm{CV} = \sigma/\mu = 12.5\%$ compared to 34.1\% (classic), 21.8\% (multiple seeds), and 51.3\% (high density). This reduced variability reflects the controlled injection geometry: walkers approach from all directions with equal probability, averaging over the anisotropic screening of individual branches. The high density case shows greatest variability because stochastic fluctuations in the local walker concentration produce intermittent bursts of rapid deposition followed by quiescent periods.

Panel~(c) of Fig.~\ref{fig:growth_dynamics} displays normalized growth curves $N(t)/N_{\mathrm{max}}$ versus normalized time $t/T$, enabling direct comparison of growth dynamics across configurations. The curves collapse reasonably well, suggesting universal growth kinetics modulo the finite-density corrections affecting Case~4. Panel~(d) of Fig.~\ref{fig:growth_dynamics} shows growth efficiency $N/R_{\mathrm{eff}}$, where the effective radius is estimated via $R_{\mathrm{eff}} = N^{1/D_f}$ from~\eqref{eq:fractal_scaling}. This quantity measures mass accumulation per unit spatial extent and remains approximately constant after initial transients, confirming self-similar growth.

Power-law fits to the growth curves yield exponents $\alpha$ in the range 1.07--1.10, substantially exceeding the theoretical prediction $\alpha = D_f/2 \approx 0.86$ from~\eqref{eq:growth_scaling}. This discrepancy arises because the theoretical scaling assumes walkers are released one at a time from infinity, whereas our simulations evolve all $N_w$ walkers simultaneously on a finite lattice with re-injection. The effective time variable in the simulation (snapshot index) does not map linearly onto the theoretical time (number of walkers released), invalidating direct comparison of growth exponents. The Ljung--Box test rejects white noise for all cases ($p < 10^{-10}$), confirming significant temporal correlations in growth rate with characteristic correlation lengths of 3--6 snapshots. The Kruskal--Wallis test yields $H = 463.7$ ($p = 3.45 \times 10^{-100}$), establishing highly significant differences in growth rate distributions across configurations.

\begin{figure}[H]
\centering
\includegraphics[width=0.8\textwidth]{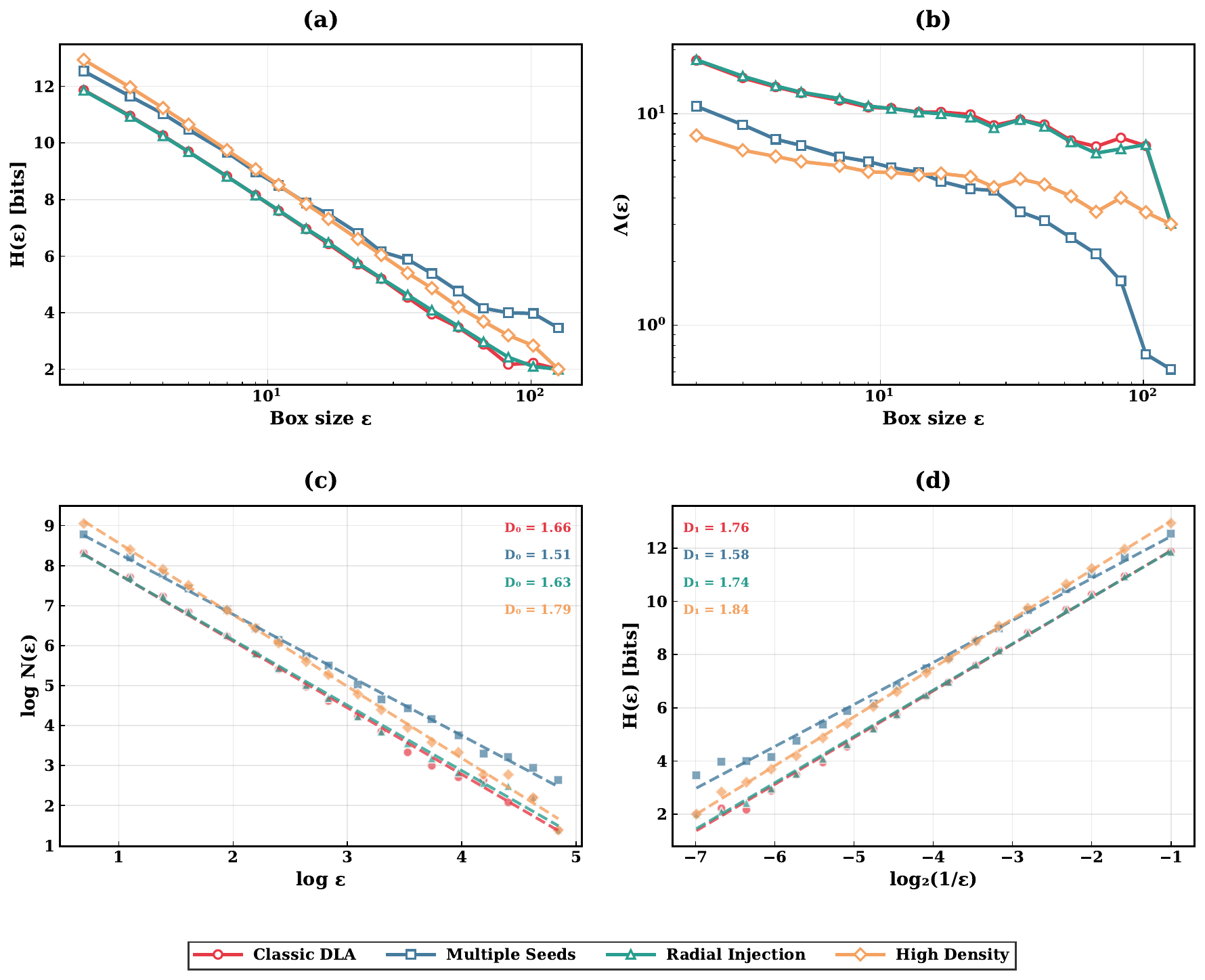}
\caption{Information-theoretic characterization of DLA aggregate structure. (a) Shannon entropy $H(\varepsilon)$ versus box size $\varepsilon$ on semi-logarithmic axes, quantifying spatial information content at each scale. (b) Lacunarity $\Lambda(\varepsilon)$ versus $\varepsilon$ on log-log axes, measuring scale-dependent heterogeneity (``gappiness''). (c) Box-counting dimension $D_0$ extraction via linear regression of $\log N(\varepsilon)$ versus $\log \varepsilon$; fitted values indicated for each case. (d) Information dimension $D_1$ extraction from $H(\varepsilon)$ versus $\log_2(1/\varepsilon)$; note consistent use of $\log_2$ matching Shannon entropy definition. For monofractal structures, $D_0 \approx D_1 \approx D_2 \approx D_f \approx 1.71$.}
\label{fig:spatial_entropy}
\end{figure}

Figure~\ref{fig:spatial_entropy} presents the information-theoretic analysis. Panel~(a) of Fig.~\ref{fig:spatial_entropy} shows Shannon entropy $H(\varepsilon)$ versus box size $\varepsilon$ on semi-logarithmic axes. All cases exhibit monotonically decreasing entropy with increasing box size, as expected: larger boxes average over more fine-scale structure, reducing the effective number of distinguishable configurations. The high density case shows highest entropy at all scales, reflecting its greater total mass and more uniform spatial distribution. The multiple seeds case shows anomalously high entropy at large scales due to the distributed cluster geometry.

Panel~(b) of Fig.~\ref{fig:spatial_entropy} displays lacunarity $\Lambda(\varepsilon)$ on log-log axes. All single-aggregate cases (1, 3, 4) exhibit similar lacunarity profiles, decreasing from $\Lambda \sim 10$ at small scales to $\Lambda \sim 1$ at large scales. This scale-dependent heterogeneity reflects the hierarchical branching structure: at fine scales, mass is concentrated along branch backbones with large empty regions between; at coarse scales, the aggregate appears more homogeneous. The multiple seeds case shows dramatically lower lacunarity at large scales ($\Lambda \sim 0.7$), reflecting the space-filling distribution of disconnected clusters.

Panels~(c) and~(d) of Fig.~\ref{fig:spatial_entropy} extract the box-counting dimension $D_0$ and information dimension $D_1$ from log-log regression. For classic DLA, we obtain $D_0 = 1.665 \pm 0.031$ and $D_1 = 1.758 \pm 0.029$; for radial injection, $D_0 = 1.634 \pm 0.028$ and $D_1 = 1.743 \pm 0.023$. These values bracket the mass-radius dimension $D_f \approx 1.71$, as expected for monofractal structures where $D_0 \leq D_1 \leq D_2$~\cite{Hentschel1983}. The small spread $D_2 - D_0 \approx 0.12$ confirms that DLA aggregates are approximately monofractal, lacking the strong multifractal character observed in some growth processes~\cite{Amitrano1986,Lee1988}. The correlation dimension $D_2 = 1.783 \pm 0.032$ (classic) and $D_2 = 1.770 \pm 0.030$ (radial) complete the generalized dimension spectrum.

The high density case yields systematically elevated dimensions: $D_0 = 1.792$, $D_1 = 1.840$, $D_2 = 1.865$. The ordering $D_0 < D_1 < D_2 < D_f$ and the small spread confirm monofractal scaling, but the absolute values reflect the more compact morphology discussed above. The multiple seeds case shows suppressed dimensions ($D_0 = 1.514$, $D_1 = 1.580$, $D_2 = 1.595$) because the box-counting procedure measures the effective dimension of the cluster distribution rather than individual aggregate structure. This distinction underscores the importance of matching analysis methods to morphological characteristics.

The overall agreement between our computed fractal dimensions and established theoretical values validates the implementation. For the two canonical configurations (classic DLA and radial injection), the mass-radius dimension deviates from $D_f = 1.71$ by less than 0.3\%, comparable to or better than published numerical studies employing similar lattice sizes~\cite{Tolman1989,Ossadnik1992}. The generalized dimensions $D_0$, $D_1$, and $D_2$ cluster within $\pm 0.05$ of $D_f$, confirming monofractal scaling consistent with theoretical expectations~\cite{Halsey1986}. The observed deviations in the high density case provide quantitative evidence for finite-density corrections to DLA universality, a regime that has received less systematic numerical investigation~\cite{Ball1985}.

\section{Conclusion}

We have demonstrated that JIT-compiled Python provides a viable pathway for high-throughput stochastic simulations, achieving statistical convergence comparable to legacy implementations without sacrificing codebase flexibility. Our analysis confirms the robustness of the standard DLA universality class ($D_f \approx 1.71$) under radial and point-source injection, yet the observed deviation in the high-density regime ($D_f \rightarrow 1.87$) serves as a cautionary note on the limits of the discrete-walker approximation. This drift likely indicates a physical crossover toward the Eden model driven by the saturation of the screening length, rather than a mere numerical artifact. While our multifractal and lacunarity analyses offer a detailed morphological fingerprint of these aggregates, a rigorous renormalization group treatment of the finite-density transition remains an open theoretical challenge. The software framework developed here offers a reliable, accessible testbed for addressing such questions in non-equilibrium statistical mechanics.

\section*{Acknowledgments}

This work was supported by the Research, Community Service and Innovation Program (FITB.PPMI-1-04-2025) and Early Career Research Scheme through the Directorate of Research and Innovation (2352/IT1.B07.1/TA.00/2025) from Bandung Institute of Technology.

\section*{Author Contributions}
S.H.S.H.: Conceptualization; Formal analysis; Investigation; Methodology; Software; Validation; Visualization; Writing – original draft; Project administration. F.R.F: Supervision; Writing – review \& editing. I.P.A.: Funding acquisition; Supervision; Writing – review \& editing. F.K.: Funding acquisition; Supervision; Writing – review \& editing. N.J.T: Supervision; Writing – review \& editing. R.S: Funding
acquisition; Supervision; Writing – review \& editing.  D.E.I.: Funding acquisition; Supervision; Writing – review \& editing. All authors reviewed and approved the final version of the
manuscript.

\section*{Open Research}

To ensure full reproducibility and transparency, all software and data associated with this study are openly available under the MIT License. The source code for the core simulation engine, \texttt{dla-ideal-solver}, is hosted on GitHub at \url{https://github.com/sandyherho/dla-ideal-solver} and is available for installation via the Python Package Index (PyPI) at \url{https://pypi.org/project/dla-ideal-solver/}. The Python scripts utilized for post-processing, statistical analysis, and figure generation are archived at \url{https://github.com/sandyherho/dla-paper-scripts}. Additionally, the complete reproducibility dataset—comprising raw simulation data, computational logs, detailed statistical outputs, and high-resolution figures—is deposited in the Open Science Framework (OSF) repository at \url{https://doi.org/10.17605/OSF.IO/AQKDP}.

\end{document}